# Frontiers of finance: Evolution and efficient markets

J. DOYNE FARMER*† AND ANDREW W. LO‡

*Prediction Company, 236 Montezuma Avenue, Santa Fe, NM 87501; and ‡Massachusetts Institute of Technology Sloan School of Management, 50 Memorial Drive, E52-432, Cambridge, MA 02142-1347

adap-org/9912001  8 Dec 1999

**ABSTRACT** In this review article, we explore several recent advances in the quantitative modeling of financial markets. We begin with the Efficient Markets Hypothesis and describe how this controversial idea has stimulated a number of new directions of research, some focusing on more elaborate mathematical models that are capable of rationalizing the empirical facts, others taking a completely different tack in rejecting rationality altogether. One of the most promising directions is to view financial markets from a biological perspective and, specifically, within an evolutionary framework in which markets, instruments, institutions, and investors interact and evolve dynamically according to the "law" of economic selection. Under this view, financial agents compete and adapt, but they do not necessarily do so in an optimal fashion. Evolutionary and ecological models of financial markets is truly a new frontier whose exploration has just begun.

If, in January, 1926, an individual invested $1 in one-month U.S. Treasury bills—one of the safest securities in the world—and continued reinvesting the proceeds month by month until December, 1996, the original investment would have grown to $14. If, on the other hand, an individual invested $1 in the S&P 500—a much riskier investment—over the same 71-year period, this investment would have grown to $1,370, a considerably larger sum. Now suppose that, each month, an individual were able to divine in advance which of these two investments would yield a higher return for that month and took advantage of this information by switching the running total of his initial $1 investment into the higher-yielding asset. What would a $1 investment in such a "perfect foresight" investment strategy become by December 1996?

The startling answer is $2,296,183,456, more than two billion dollars!§ Despite the fact that perfect foresight in financial markets is impossible, this example suggests that even a modest ability to forecast financial asset returns may be handsomely rewarded. Of course, there are considerable risks involved. As recent financial calamities such as that of Long Term Capital Management have made very clear, it is also possible to lose very large sums of money. Financial markets are very difficult to predict; otherwise, we would all be rich. One of the central questions, indeed, perhaps the most central question in finance, is, "Under what circumstances is prediction possible at all?"

In this review article, we explore this issue in light of recent advances in the quantitative modeling of financial markets. The potent combination of breakthroughs in financial technology and computational speed and efficiency is creating an exciting renaissance in financial research, both inside and outside the halls of academia. It is impossible for us to provide a complete survey of these developments here; instead, we focus on the beginnings of a new research direction that the emerging fields of computational finance and financial engineering may be heading toward— evolutionary and ecological models of financial markets—and how these new perspectives may be changing fundamental views about market prediction.

Our starting point is the "Efficient Markets Hypothesis" (EMH), a powerful idea that can be traced back to Paul Samuelson (1), whose contribution is neatly summarized by the title of his article: "Proof that Properly Anticipated Prices Fluctuate Randomly." In an informationally efficient market, price changes must be unforecastable if they are properly anticipated, i.e., if they fully incorporate the expectations and information of all market participants.

This concept of informational efficiency has a wonderfully counterintuitive and seemingly contradictory flavor to it: the more efficient the market, the more random the sequence of price changes generated by such a market must be, and the most efficient market of all is one in which price changes are completely random and unpredictable. This, of course, is not an accident of nature but is the direct outcome of many active participants attempting to profit from their information. Unable to curtail their greed, an army of investors aggressively pounce on even the smallest informational advantages at their disposal, and, in doing so, they incorporate their information into market prices and quickly eliminate the profit opportunities that gave rise to their actions. If this occurs instantaneously, which it must in an idealized world of "frictionless" markets and costless trading, then prices must always fully reflect all available information, and no profits can be garnered from information-based trading (because such profits have already been captured).

But one of the central tenets of modern financial economics is the necessity of some trade-off between risk and expected return. If a security's expected price change is positive, it may be just the reward needed to attract investors to hold the asset and bear the corresponding risks. Indeed, if an investor is sufficiently risk averse, he might gladly pay to avoid holding a security that has unforecastable returns. In such a world, prices do not need to be perfectly random, even if markets are operating efficiently and rationally.

Indeed, several statistical studies have made it clear that prices are, in fact, not completely random (see, for example, ref. 2). Economists disagree on whether this represents a violation of efficient markets. Similarly, the sustained profits of some investment companies and certain high-profile portfolio managers seem to challenge the very foundations of market rationality and efficiency. Are they just lucky? Are they merely receiving appropriate compensation for risk? Or are markets simply inefficient? Such questions have proved to be difficult to answer and remain controversial.

---

Abbreviation: EMH, Efficient Markets Hypothesis.
†To whom reprint requests should be addressed. E-mail: jdf@santafe.edu.
§We are grateful to Bob Merton (Harvard University Business School) for this example.

One of the reasons for this state of affairs is the fact that the EMH, by itself, is not a well posed and empirically refutable hypothesis. To make it operational, one must specify additional structure: e.g., investors' preferences, information structure, etc. But then a test of the EMH becomes a test of several auxiliary hypotheses as well, and a rejection of such a joint hypothesis tells us little about which aspect of the joint hypothesis is inconsistent with the data. Moreover, new statistical tests designed to distinguish among them will no doubt require auxiliary hypotheses of their own that, in turn, may be questioned. The hypothesis that investors are fully rational agents that instantaneously and correctly process all available information is clearly unrealistic—rationality is difficult to define, human behavior is often unpredictable, information can be difficult to interpret, technology and institutions change constantly, and there are significant "frictional" costs to gathering and processing information and transacting. But how can we take all of the complexities of the real world into account?

One new direction is to treat the EMH as an idealization that provides a useful reference point. For example, one can ask about the relative efficiency of markets with respect to each other: e.g., futures vs. spot markets, auction vs. dealer markets, etc. The advantages of the concept of relative efficiency, as opposed to the all-or-nothing notion of absolute efficiency, are easy to spot by way of an analogy to the concept of efficiency as used in physics. Heat engines can be given an efficiency rating based on the fraction of available energy that they convert into useful work. A refrigerator with an efficiency of 40% might be considered quite good, and a buyer would prefer this to one with an efficiency of only 35%. No one would ever expect 100% efficiency. The best measure of the relative efficiency of financial markets, relative to each other, is a topic on the frontiers of research in finance.

Another point of view is to extend the definition of efficient markets so that consistent profits are possible to those who acquire a competitive advantage. The motivation for this becomes apparent from applying the classical version of the EMH to a nonfinancial context, such as a biotechnology firm attempting to develop a vaccine for the AIDS virus. If the market for biotechnology is efficient in the classical EMH sense, such a vaccine can never be developed—if it could, someone would have already done it! This is clearly an absurd conclusion because it ignores the challenges and gestation lags of research and development in biotechnology. If a pharmaceutical company does succeed in developing such a vaccine, the profits earned might be measured in the billions of dollars—would this be considered "excess" profits or simply an appropriate economic reward for competence and innovation? Financial markets should not be different in principle, only in degree. The profits that accrue to an investment professional need not be a market inefficiency but may simply be the fair reward for unusual skill, extraordinary effort, or for breakthroughs in financial technology.

What, then, can we conclude about the EMH? Amazingly, there is still no consensus among financial economists. Despite the many advances in the statistical analysis, databases, and theoretical models surrounding the EMH, the main effect has been to harden the resolve of the proponents on each side.

However, the controversy surrounding the EMH has stimulated a number of new directions of research, some focusing on more elaborate mathematical models that are capable of rationalizing the empirical facts, others taking a completely different tack in rejecting rationality altogether. We think one of the most promising directions is to view financial markets from a biological perspective and, specifically, within an evolutionary framework in which markets, instruments, institutions, and investors interact and evolve dynamically according to the "law" of economic selection. Under this view, financial agents compete and adapt, but they do not necessarily do so in an optimal fashion.

The desire to build financial theories based on more realistic assumptions has led to several new strands of literature, including psychological approaches to risk-taking behavior (3–5), evolutionary game theory (6), and agent-based modeling of financial markets (ref. 7; N. Chan, B. LeBaron, A.L., and T. Poggio, unpublished work). Although substantially different in methods and style, these emerging subfields are all directed at new interpretations of the EMH. In particular, psychological models of financial markets focus on the manner in which human psychology influences the economic decision-making process as an explanation of apparent departures from rationality. Evolutionary game theory studies the evolution and steady-state equilibria of populations of competing strategies in highly idealized settings. Agent-based models are meant to capture complex learning behavior and dynamics in financial markets by using more realistic markets, strategies, and information structures.

For example, in one agent-based model of the financial markets (J.D.F., unpublished work), the market is modeled by using a nonequilibrium market mechanism whose simplicity makes it possible to obtain analytic results while maintaining a plausible degree of realism. Market participants are treated as computational entities that employ strategies based on limited information. Through their (sometimes suboptimal) actions, they make profits or losses. Profitable strategies accumulate capital with the passage of time, and unprofitable strategies lose money and may eventually disappear. A financial market can thus be viewed as a coevolving ecology of trading strategies. The strategy is analogous to a biological species, and the total capital deployed by agents following a given strategy is analogous to the population of that species. The creation of new strategies may alter the profitability of preexisting strategies, in some cases replacing them or driving them extinct.

Although agent-based models are still in their infancy, the simulations and related theory have already demonstrated an ability to understand many aspects of financial markets. Several studies indicate that, as the population of strategies evolve, the market tends to become more efficient, but this is far from the perfect efficiency of the classical EMH. Prices fluctuate in time with internal dynamics caused by the interaction of diverse trading strategies. Prices do not necessarily reflect "true values"; if we view the market as a machine whose job is to set prices properly, the inefficiency of this machine can be substantial. Patterns in the price tend to disappear as agents evolve profitable strategies to exploit them, but this occurs only over an extended period of time, during which substantial profits may be accumulated and new patterns may appear.

Thomas Malthus and Adam Smith—two of the forefathers of modern economics—were both cited by Darwin as inspirations for the principle of natural selection, and analogies between economics and biology have been discussed for more than a century. However, a quantitative foundation for this approach has been slow to develop. Recent research in finance suggests that this is about to change (5). Although there are obvious differences between evolution in biological systems and evolution in financial systems, there are also many similarities. The theory of evolution may prove to be as powerful an idea in finance as it has been in biology. There is no lack of quantitative data, and there are many opportunities for biological principles to be applied to financial interactions—after all, financial institutions are uniquely human inventions that provide an adaptive advantage to our species. This is truly a new frontier whose exploration has just begun.

We thank the participants of the National Academy of Sciences 1998 Frontiers of Science Symposium for valuable discussions and comments. This research was partially supported by the Massachusetts Institute of Technology Laboratory for Financial Engineering and the National Science Foundation (Grant SBR-9709976).

1. Samuelson, P. (1965) *Industrial Management Rev.* **6,** 41–49.
2. Lo, A. & MacKinlay, C. (1999) *A Non-Random Walk Down Wall Street* (Princeton Univ. Press, Princeton).
3. Kahneman, D. & Tversky, A. (1979) *Econometrica* **47,** 263–291.
4. Thaler, R. (1993) *Advances in Behavioral Finance* (Russell Sage Foundation, New York).
5. Lo, A. (1999) *Financial Analysts J.* **55,** 13–26.
6. Friedman, D. (1991) *Econometrica* **59,** 637–666.
7. Arthur, W. B., Holland, J., LeBaron, B., Palmer, R. & Taylor, P. (1997) in *The Economy As An Evolving Complex System*, eds. Arthur, W. B., Durlauf, S. & Lane, D. (Addison–Wesley, Reading, MA), Vol. II.